\documentclass{aa}

\usepackage[varg]{txfonts}
\usepackage{amsmath}
\usepackage{amssymb}
\usepackage{wrapfig}
\usepackage{graphicx}
\usepackage{natbib}
\usepackage{caption}[2008/08/24]
\usepackage{txfonts}

\title{Evolution of a magnetic field in a differentially rotating radiative zone}
\author{M. Gaurat \inst{\ref{irap1},\ref{irap2}}
\and L. Jouve\inst{\ref{irap1},\ref{irap2}}
\and F. Ligni\`eres\inst{\ref{irap1},\ref{irap2}}
\and T. Gastine\inst{\ref{mps}}}

\institute{Universit\'e de Toulouse, UPS-OMP, Institut de Recherche en Astrophysique et Plan\'etologie, 31028 Toulouse Cedex 4, France\\
email:[mathieu.gaurat;laurene.jouve;francois.lignieres]@irap.omp.eu\label{irap1}
\and CNRS, Institut de Recherche en Astrophysique et Plan\'etologie, 14 avenue \'Edouard Belin, 31400 Toulouse, France\label{irap2}
\and Max-Planck-Institut für Sonnensystemforschung, Justus-von-Liebig-Weg 3, 37077 Göttingen, Germany\\
e-mail:gastine@mps.mpg.de\label{mps}}
\date{Received /
Accepted}

\abstract
{Recent spectropolarimetric surveys of main-sequence intermediate-mass stars have exhibited a dichotomy in the distribution of the observed magnetic field between the kG dipoles of Ap/Bp stars and the sub-Gauss magnetism of Vega and Sirius.} %One thinks that this dichotomy could be the result of an instability of the intermediate-mass stars magnetic field.}
{We would like to test whether this dichotomy is linked to the stability versus instability of large-scale magnetic configurations in differentially rotating radiative zones.}
{We computed the axisymmetric magnetic field obtained from the evolution of a dipolar field threading a differentially rotating shell.
A full parameter study including various density profiles and initial and boundary conditions was performed with a 2D numerical code.
We then focused on the ratio between the toroidal and poloidal components of the magnetic field and discuss the stability of the configurations dominated by the toroidal component using local stability criteria and insights from recent 3D numerical simulations.}
{The numerical results and a simple model show that the ratio between the toroidal and the poloidal magnetic fields is highest after an Alfv\'en crossing time of the initial poloidal field.
For high density contrasts, this ratio converges towards an asymptotic value that can thus be extrapolated to realistic stellar cases.
We then consider the stability of the magnetic configurations to non-axisymmetric perturbations and find that configurations dominated by the toroidal component are likely to be unstable if the shear strength is significantly higher than the poloidal Alfv\'en frequency.
An expression for the critical poloidal field below which magnetic fields are likely to be unstable is found and is compared to the lower bound of Ap/Bp magnetic fields.}
{}


\begin{document}

\maketitle

\section{Introduction}
In the past decades, the study of the interplay between magnetic fields and differential rotation in radiative zones has mainly been driven by constraints on stellar rotation. 
This interaction has in particular been invoked to explain the flat rotation profile of the solar radiative zone \citep[see for example][]{MW87, CM93} or more recently, the slow rotation of the core of subgiants and giants revealed by asteroseismology \citep{deheuvels2014,cantiello2014}.

\raggedbottom
The evolution of the angular momentum in the presence of a magnetic field has mainly been investigated assuming axisymmetry and neglecting meridional circulation and turbulence.
This was the case of the numerical simulations by \cite{CM92} that considered the spin-up of a radiative stellar zone threaded by an initial poloidal field.
The authors pointed out the role of the field geometry, more specifically, the existence of field lines anchored in the core, in reaching or failing to reach a solid-body rotation as a final state.
They determined the timescale necessary to reach this state.
This timescale is controlled by the damping of the Alfv\'en waves that are excited by the back-reaction of the Lorentz force that follows the winding-up of the poloidal field by the differential rotation.
\cite{CM93}, \cite{RK96}, and \cite{spada2010} studied the spin-down of the radiative zone of the Sun under the same physical assumptions.
The effect of meridional flows has been addressed in \cite{MMT88}, who restricted their analysis to the axisymmetric case. 
The impact of non-axisymmetric field components was then studied by \cite{MMT90}, \cite{moss92} and more recently by \cite{WG2015}.
It was shown that a state of solid-body rotation is reached on a timescale shorter than the global diffusive time, regardless of the existence of field lines anchored in the core. 
However, if the field strength is small compared to the differential rotation strength, the misaligned magnetic field can be axisymmetrised before solid-body rotation is reached.

These studies did not include the effect of possible non-axisymmetric instabilities of the magnetic configurations.
However, on the one hand, such instabilities are expected if the winding-up of the poloidal field by differential rotation produces magnetic configurations that are dominated by the toroidal component.
Purely toroidal field configurations can indeed be unstable to various kinds of instabilities, such as the Tayler instability, the magneto-rotational instability (MRI), or the buoyancy instability \citep[see for a review][]{spruit99}.
On the other hand, magnetic fields with toroidal and poloidal components of the same order of magnitude can remain stable, as was found in recent numerical simulations \citep[e.g.][]{BN2006,braithwaite2007}.
For the angular momentum transport, the occurrence of such instabilities is a crucial issue since the development of a non-axisymmetric instability could profoundly modify the redistribution of angular momentum.
Estimates of the transport resulting from the Tayler instability have been proposed in \cite{spruit2002} on phenomenological grounds, while recent numerical simulations by \cite{rudiger2015}
quantified the transport induced by the so-called azimuthal-MRI in the simplified setup of a fluid of constant density confined between two rotating cylinders.

Beyond the problem of angular momentum transport, non-axisymmetric instabilities could also affect the value of the magnetic field measured at the surface of stars.
Notably, spectropolarimetry is an observational technique that allows measuring the line-of-sight component of the magnetic field vector integrated over the whole visible surface.
This averaged field could be strongly decreased as one goes from a well-organized large scale magnetic field to a destabilized configuration where opposite polarities cancel each other, especially if the typical length scale left by the instability is much smaller than the stellar radius. 
This effect has been invoked to explain the magnetic dichotomy observed among intermediate-mass stars between the strong mostly dipolar Ap/Bp magnetic fields and the ultra-weak magnetic fields of other intermediate-mass stars \citep{auriere2007,lignieres2014}.
In this scenario, the observed lower bound of Ap/Bp magnetic fields ($\sim 300$ Gauss for the dipolar strength) corresponds to the limit between stable and unstable magnetic fields.
On phenomenological grounds, \cite{auriere2007} approximated the maximum toroidal field produced by the winding-up of the poloidal field to $B_{\varphi} = R \, \Omega\sqrt{4\pi\rho}$ (where $R$ is the radius of the star, $\Omega$ is the rotation rate and $\rho$ is the density) and assumed that the magnetic configuration is unstable as soon as the toroidal field dominates the poloidal field at the stellar surface.
They thus derived an order-of-magnitude estimate of the critical field $B_c = R \, \Omega\sqrt{4\pi\rho}$ that separates stable and unstable magnetic fields.
This estimate appears to match the observed value for a typical Ap/Bp star.

Our ultimate goal is to investigate the occurrence and the effect of non-axisymmetric instabilities of the magnetic field induced by the winding-up of a poloidal field in a differentially rotating stellar radiative zone.
In the present paper, we use axisymmetric simulations to explore the different magnetic configurations obtained for various differential rotation profiles, density stratifications, and boundary conditions.
In this systematic parameter study, we focus on the ratio between the toroidal and poloidal magnetic field as a key parameter governing the stability of the magnetic configuration. 
We discuss the occurrence of instability in fields that are dominated by their toroidal component, for which we use local stability criteria for purely toroidal fields as well as results from recent 3D numerical simulations by \citep{jouve2015} performed for particular configurations.
The advantage of these 2D axisymmetric numerical simulations over 3D simulations is to give access to a much broader parameter range and in particular to explore asymptotic behaviours in the regimes of low magnetic diffusivities and viscosities and large density stratifications that are unaccessible to 3D models.

In Sect. \ref{math}, the mathematical formulation of our problem is described. We set down the basic assumptions that we adopted and then introduce the 
equations governing the joint evolution of the differential rotation and the magnetic field.
In Sect. \ref{results}, the results of the numerical simulations are presented. 
In Sect. \ref{models}, simple models that describe the evolution of the toroidal to poloidal field ratio are presented and compared to the numerical results.
In Sect. \ref{instability}, we discuss the stability of the magnetic configurations obtained by the simulations.

\section{Mathematical formulation}
\label{math}

\subsection{Simplifying assumptions and governing equations}
We consider the axisymmetric evolution of an initially poloidal field submitted to differential rotation in a spherical shell.
Beyond the assumption that the flow remains axisymmetric over time, we also neglect the meridional circulation. 
The dynamics is therefore not affected by buoyancy effects and is only governed by the induction and the azimuthal momentum equations.
The stellar equilibrium structure comes into play through the density stratification.
We also consider timescales shorter than the magnetic diffusion time on a radius length scale. 
Under the assumptions of axisymmetry and without meridional circulation, the initial poloidal field only evolves
through Ohmic dissipation and is thus considered as constant in time.

The governing equations are thus
\begin{align}
\label{ind}
 \frac{{\partial}B_\varphi}{{\partial}t}&=r\sin\theta \, (\vec{B_p}\cdot\vec{\nabla}) \, \Omega+\eta\left(\Delta-\frac{1}{r^2\sin^2\theta}\right)B_\varphi\ \ \ ,\\
\begin{split}
 \label{ns}
 {\rho} \, r\sin\theta \, \frac{{\partial}\Omega}{{\partial}t} &= \frac{1}{4{\pi}r\sin\theta}(\vec{B_p}\cdot\vec{\nabla})(r\sin\theta \, {B_\varphi}) \\
 &+\mu\left(\Delta-\frac{1}{r^2\sin^2\theta}\right)(r\sin\theta \, \Omega)\ \ \ ,
\end{split}
\end{align}
where the rotation rate $\Omega(r,\theta,t)$ is related to the azimuthal velocity by $u_\varphi(r,\theta,t)=r\sin\theta \, \Omega(r,\theta,t)$
and the magnetic field reads
\begin{equation}
\label{B}
 \vec{B}(r,\theta,t)=\vec{B_p}(r,\theta)+B_\varphi(r,\theta,t) \, \vec{e_\varphi}\ \ \ ,
\end{equation}
$B_{\varphi}(r,\theta,t)$ being its azimuthal component and $\vec{B_p}(r,\theta)$, the time-independent poloidal component.
In these equations, the magnetic diffusivity $\eta$ and the dynamic viscosity $\mu$ are uniform, while the density $\rho$ is given by a polytropic model of a star with a polytropic index $n=3$ \citep{rieutord2005}.

\subsection{From three to two dimensionless numbers}

A possible and straightforward nondimensionalization of these equations is obtained by taking the stellar radius $R$ as the reference length scale, $1/\Omega_0$
as the reference timescale (where $\Omega_0$ is the surface rotation rate at the equator), $B_0$ the surface poloidal magnetic field taken at the equator as a magnetic field unit, and $\rho_0$ the surface density as a density unit.
The dynamical Eqs. (\ref{ind}) and (\ref{ns}) are then controlled by three different dimensionless numbers, namely the Reynolds number $Re ={\Omega_0} \, R^2/\nu_0$, where $\nu_0=\mu/\rho_0$ is the kinematic viscosity at the surface, the Elsasser number $\Lambda=B_{0}^2/(4\pi\rho_0 \, \Omega_0 \, \eta)$, and the magnetic Prandtl number $P_m=\nu_0/\eta$.
However, it is also possible to reduce the number of dimensionless parameters to two by instead employing the Alfv\'en timescale $t_{Ap}=R\sqrt{4\pi\rho_0}/B_0$ as the reference timescale and $R \, \Omega_{0}\sqrt{4\pi\rho_0}$ as a magnetic field unit for the toroidal component.
In addition to the reduction of the number of dimensionless parameters allowed by this choice, it is motivated by the fact that $B_\varphi$ is generated by the $\Omega$-effect and hence related to $\Omega_0$ in a differentially rotating radiative zone.
$B_0$ remains the poloidal magnetic field unit and $\Omega_0$ the angular rate unit for this nondimensionalization.
The ratio of the toroidal to poloidal magnetic fields is then expressed as
\begin{equation}
\label{bphibp}
 \frac{B_\varphi}{B_p}=\frac{\widetilde{B_\varphi}}{\widetilde{B_p}} \, \frac{t_{Ap}}{t_\Omega}\ \ \ ,
\end{equation}
where $t_\Omega=1/\Omega_0$, and where tildes indicate dimensionless quantities. 
With this nondimensionalization, Eqs. (\ref{ind}) and (\ref{ns}) read
\begin{align}
\label{effetomega}
\frac{{\partial}\widetilde{B_\varphi}}{{\partial}\widetilde{t}} &= \widetilde{r}\sin\theta \, (\widetilde{\vec{B_p}}\cdot\widetilde{\vec{\nabla}}) \, \widetilde{\Omega}+\frac{1}{L_u}\left(\widetilde{\Delta}-\frac{1}{\widetilde{r}^2\sin^2\theta}\right)\widetilde{B_\varphi}\ \ \ ,\\
\begin{split}
\label{retroaction}
 {\widetilde{\rho}} \, \, \widetilde{r}\sin\theta \, \frac{{\partial}\widetilde{\Omega}}{{\partial}\widetilde{t}} &= \frac{1}{\widetilde{r}\sin\theta} \, (\widetilde{\vec{B_p}}\cdot\widetilde{\vec{\nabla}})(\widetilde{r}\sin\theta \, {\widetilde{B_\varphi}})\\
 &+\frac{P_m}{L_u}\left(\widetilde{\Delta}-\frac{1}{\widetilde{r}^2\sin^2\theta}\right)(\widetilde{r}\sin\theta \, \widetilde{\Omega})\ \ \ .
\end{split}
\end{align}
$L_u$, the Lundquist number, is defined as follows:
\begin{equation}
\label{lu}
  L_u=\frac{t_\eta}{t_{Ap}}=\frac{R \, B_0}{\eta\sqrt{4\pi\rho_0}}\ \ \ .
\end{equation}
This reduction from three to two dimensionless numbers greatly facilitates the exploration of the parameter space.

\subsection{Initial and boundary conditions}

\begin{figure}[!h]
\begin{center}
 \includegraphics[scale=0.4]{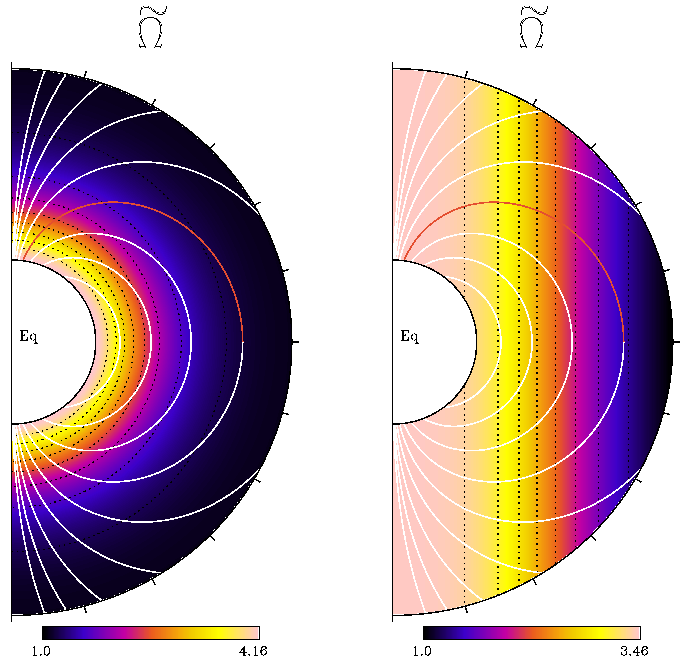}
 \caption{\small{Meridional cut of $\widetilde{\Omega}$ for the radial profile defined by Eq. \ref{ci23} (left) and for the cylindrical profile defined by Eq. \ref{ci30} (right). 
 The dotted black lines represent the isocontours of $\widetilde{\Omega}$. The white lines represent the poloidal magnetic field lines and the red line the one on which Alfvén waves propagate the slowest.
 The radius at the inner boundary is $r=R_c=0.3 \, R$}}
 \label{bp_omega}
\end{center}
\end{figure}

The initial magnetic field is a dipole (see white lines in Fig. \ref{bp_omega}):
\begin{align}
\label{Bp}
 \vec{B_p}(r,\theta,t=0) &= \frac{2\cos\theta \, B_0 \, R^3}{r^3} \, \vec{e_r}+\frac{\sin\theta \, B_0 \, R^3}{r^3} \, \vec{e_\theta}\ \ \ ,\\
 \label{Bphi}
B_\varphi(r,\theta,t=0) &= 0\ \ \ .
\end{align}
As stated above, the poloidal component is time-independent, while the toroidal component will evolve in time.
The boundary conditions are a perfect conductor at the bottom boundary and an insulating medium at the outer radius. This translates into
\begin{equation}
 \left|\frac{\partial{(rB_{\varphi}(r,\theta,t))}}{\partial{r}}\right|_{r=R_c,\theta,t}=0\phantom{aa}\text{and}\phantom{aa}B_\varphi(r=R,\theta,t)=0\ \ \ ,
\end{equation}
where $R_c$ is the radius of the inner boundary. 
We here explore different initial and boundary conditions for the rotation.
Three different radial differential rotations and two cylindrical profiles were considered:
\begin{equation}
\label{ci22}
 \Omega(r)=\protect\Omega_0 \, \frac{R}{r}\ \ \ ,
\end{equation}
\begin{equation}
\label{ci25}
 \Omega(r)=\protect\Omega_0 \, \frac{R^2}{r^2}\ \ \ ,
\end{equation}
\begin{equation}
\label{ci23}
 \Omega(r)=\protect\Omega_0\frac{1-c_1 \, (r-R_c)^2 \, R/r^3-c_2 \, (r-R_c)^2/r^2}{1-(c_1+c_2)(1-R_c/R)^2}\ \ \ ,
\end{equation}
\begin{align}
\label{ci26}
 \Omega(\varpi)&=\Omega_0\sqrt{\protect\frac{2}{1+(\varpi/R)^4}}\ \ \ ,\\
\label{ci30}
 \Omega(\varpi)&=\Omega_0\sqrt{\protect\frac{12}{1+11(\varpi/R)^4}}\ \ \ ,
\end{align}

where $\varpi=r\sin\theta$ is the distance from the rotation axis.
Two types of boundary conditions are used at $r=R_c$, namely a stress-free or a fixed rotation, while a stress-free boundary condition is always used at $r=R$. 
These different cases are summarized in Tables \ref{table_ci} and \ref{table_bc} and illustrated in Fig. \ref{bp_omega} which displays a meridional cut of a radial (left frame) and a cylindrical (right frame) differential rotation profile.

\begin{figure}[!h]
\begin{center}
 \includegraphics[scale=0.47,trim= 0cm 0cm 0cm 0cm]{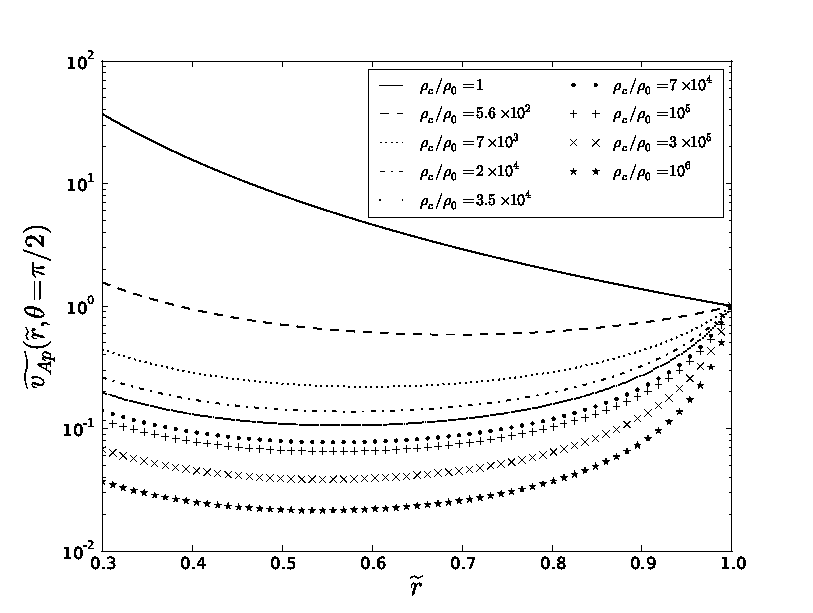}
 \caption{\small{$\widetilde{v_{Ap}}$ at the equator as a function of $\widetilde{r}$ for different density contrasts $\rho_c/\rho_0$.
 A $n=3$ polytropic profile was cut at different radii to produce these various $\rho_c/\rho_0$.}}
 \label{va}
\end{center}
\end{figure}

From our reference density profile corresponding to an $n=3$ polytrope, we create different profiles with ratios $\rho_c/\rho_0$ ranging from $1$ to $10^6$ (where $\rho_c$ is the density at the radius $r=R_c$) by cutting the polytropic profile at different radii.
The created profiles are then expanded on the whole computational domain.
Figure \ref{va} shows the radial dependence of the dimensionless Alfv\'en velocity $\widetilde{v_{Ap}}=\widetilde{B_p}/\sqrt{\widetilde{\rho}}$ for the different density profiles used in our simulations.
The corresponding time for an Alfvén wave to propagate from $r=R_c$ to $r=R$ along the equator would vary from $0.25 \, t_{Ap}$ for $\rho_c/\rho_0=1$ to $22 \, t_{Ap}$ for $\rho_c/\rho_0=10^6$.

\subsection{Numerical model}
To solve Eqs. (\ref{effetomega}) and (\ref{retroaction}) with the boundary and initial conditions, we used the two-dimensional axisymmetric code STELEM \citep{CM92,jouve2008}. 
This code uses a finite-element method in space and a third-order Runge-Kutta scheme in time.
The computational domain is limited to the annulus $r\in[R_c,R]$, $\theta\in[0,\pi]$, where $R_c=0.3 \, R$ was chosen for all our simulations.
$P_m$ was fixed to $1$ in all our simulations while the Lundquist number $L_u$ is varied from $10$ to $10^5$.
\begin{figure*}
\begin{center}
 \includegraphics[scale=0.24,trim= 0cm 0cm 0cm 0cm]{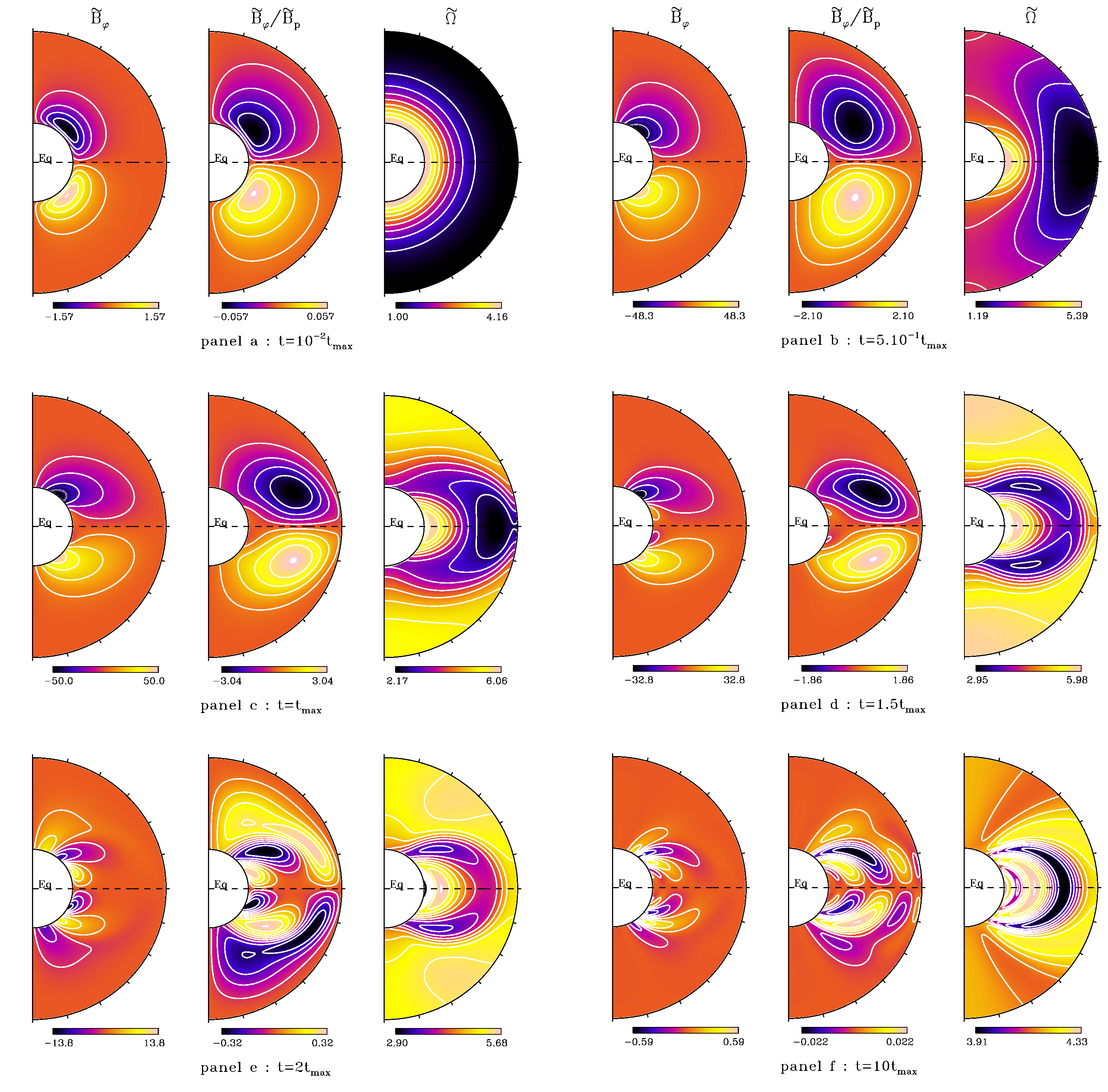}
 \caption{\small{The reference case: Meridional cut of $\widetilde{B_\varphi}$, $\widetilde{B_\varphi}/\widetilde{B_p}$ and $\widetilde{\Omega}$ for different times. $t_{max}$, the time at which $B_\varphi/B_p$ is maximal, is equal to $1.85 \, t_{Ap}$ in this simulation. The colourbar at the bottom of each panel indicates the minimal and maximal values of each variable.
 The parameters of this simulation are $L_u=10^2$, a density contrast of $\rho_c/\rho_0=7\times10^3$, an initial radial differential rotation given by Eq. (\ref{ci23}) and a fixed rotation at the inner radius.}}
 \label{image2d}
\end{center}
\end{figure*}
Depending on the value of $L_u$ and on the boundary and initial conditions, different spatial and temporal resolutions are required. 
We used a spatial mesh ranging from $N_r\times{N_\theta}=128\times128$ to $512\times512$ nodes depending on the cases.
The integration time was adjusted to satisfy the Courant-Friedrichs-Lewy criterion. It varies from $dt=10^{-3} \, t_{Ap}$ to $dt=10^{-8} \, t_{Ap}$.

With the initial conditions used in this study, our problem has an equatorial symmetry. This is clearly visible in the meridional cut of Fig. \ref{bp_omega} for example.
However, for comparison to future studies with more complex symmetry, we chose to include both hemispheres in our computations and always represent the full meridional plane.

We note that we do not present any results for $L_u>245$ with a stress-free boundary condition at $r=R_c$. In fact, with this boundary condition there is a Hartmann boundary layer at the inner radius.
The thickness $\delta$ of this boundary layer is known to be \citep{dormy98}
\begin{equation}
\label{delta}
 \delta=\frac{R \, B_0}{\vec{B_p}\cdot\vec{n}}\frac{\sqrt{P_m}}{L_u}\ \ \ ,
\end{equation}
where $\vec{n}$ is the unit vector normal to the boundary.
The number of mesh points is increased according to Eq. (\ref{delta}) to ensure that the Hartmann layer is much larger than the size of the numerical grid. 
However, for large $L_u$, this imposes prohibitive restrictions on the time step.

\section{Description of the results}
\label{results}
The evolution of the magnetic field and the rotation rate is first qualitatively described pointing out the successive phases and the associated physical mechanisms. 
We then focus on the ratio between the toroidal and the poloidal magnetic field as a key parameter in the study of the stability of a magnetic field. 
The influence of the Lundquist number, the density contrast, the initial and the boundary conditions on the highest value reached by this ratio is investigated in detail. 

\subsection{Evolution of the magnetic field and the differential rotation in a typical case}
\label{case1}

Figure \ref{image2d} shows the evolution of the toroidal magnetic field $B_\varphi$, the ratio between $B_\varphi$ and the poloidal magnetic field $B_p$, and the rotation rate $\Omega$ for a typical case (hereafter called the reference case). 
The parameters of this simulation are $L_u=10^2$, a density contrast $\rho_c/\rho_0=7\times10^3$, an initial radial differential rotation given by Eq. (\ref{ci23}) and a fixed rotation at the inner radius.
Different phases in the evolution of $B_\varphi$ and $\Omega$ can be distinguished. The first is the winding-up phase, where the shearing of the poloidal field by the differential rotation, the so-called $\Omega$-effect, generates $B_\varphi$ without any back-reaction on the differential rotation (panel $a$). 
Since $B_\varphi$ is generated by the term $\vec{B_p}\cdot\vec\nabla\Omega$, the fact that $B_\varphi$ is antisymmetric with respect to the equator directly relates to the properties of symmetry of $\Omega$ and $\vec{B_p}$.
After this phase, the Lorentz force back-reacts on the differential rotation, leading to the propagation of Alfv\'en waves along the poloidal magnetic field lines in both directions. 
In this simulation, we mainly observe an outward propagation from the internal radius because the initial perturbation of the system is concentrated close to the bottom boundary. 

In panel $d$, a region of opposite sign is visible on $B_\varphi$ close to the core. It is due to the reflection of Alfv\'en waves on the equator, where conditions on $B_\varphi$ and $\Omega$ allow for such reflections.
Indeed, the ability for a boundary to reflect Alfv\'en waves depends on the characteristics of $B_\varphi$ and $\Omega$ at this boundary. 
In our simulations, the equator behaves as a boundary where $B_\varphi=0$ (insulating condition) and $\partial\Omega/\partial\theta=0$ (stress-free condition).
This pair of boundary conditions induces a perfect reflection of Alfv\'en waves at the equator \citep{schaeffer2012}.
The subsequent evolution is characterized by the propagation and the reflection of the waves either at the equator, at the surface, or at the bottom boundary.

The variable Alfv\'en velocity and the different distances that the Alfv\'en waves have to travel before reflexion lead to a phase shift between waves on neighbouring magnetic field lines. 
As we can see in panel $e$, strong gradients of $B_\varphi$ and $\Omega$ are then created between these waves. The effect of magnetic and viscous diffusion is thus increased. 
This is the phase-mixing mechanism as described in \cite{ionson78}, \cite{HP83}, or \cite{parker91}.
The oscillations of $B_\varphi$ then decrease in amplitude, as shown in panel $f$. 
After this dissipative phase controlled by the phase-mixing mechanism, the system evolves towards a steady state known as Ferraro's law of isorotation \citep{ferraro37}, in which $\Omega$ is constant along poloidal field lines. 
In our case, the only possible steady state compatible with the boundary conditions and Ferraro's law is uniform rotation. 
This state of uniform rotation is reached on timescales of the order of a few diffusive timescales which are much longer than the timescales we are interested in here.
It is thus not shown in Fig. \ref{image2d}.

\begin{figure}[!h]
\begin{center}
 \includegraphics[scale=0.47,trim= 0.5cm 0cm 0cm 5mm]{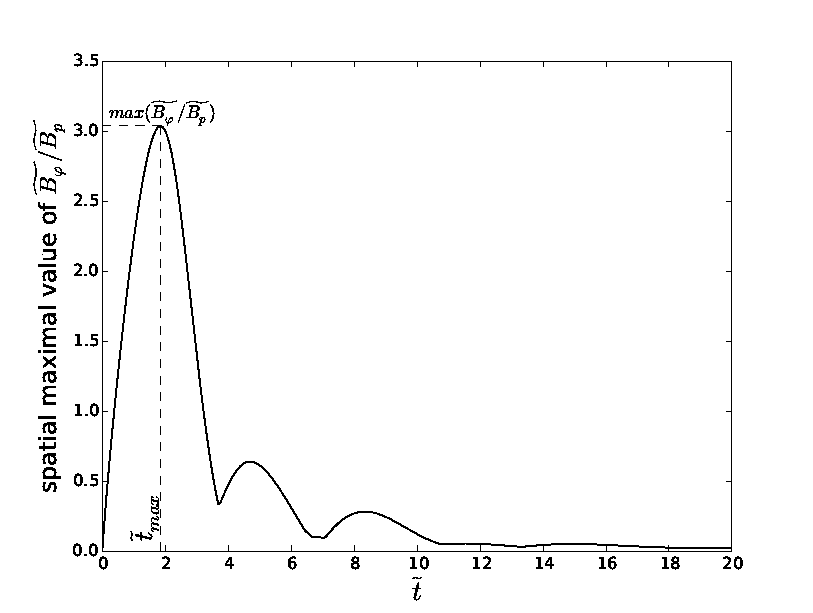}
 \caption{\small{Evolution of the spatial maximal value of $\widetilde{B_\varphi}/\widetilde{B_p}$ for the reference case. The horizontal dashed line indicates $max(\widetilde{B_\varphi}/\widetilde{B_p})$ and the vertical dashed line indicates $\widetilde{t}_{max}$.}}
 \label{max}
\end{center}
\end{figure}

\subsection{Evolution of the ratio between the toroidal and the poloidal magnetic field}
\label{evolution}

\begin{figure*}
\begin{center}
 \includegraphics[scale=0.43,trim = 10cm 0cm 10cm 0cm]{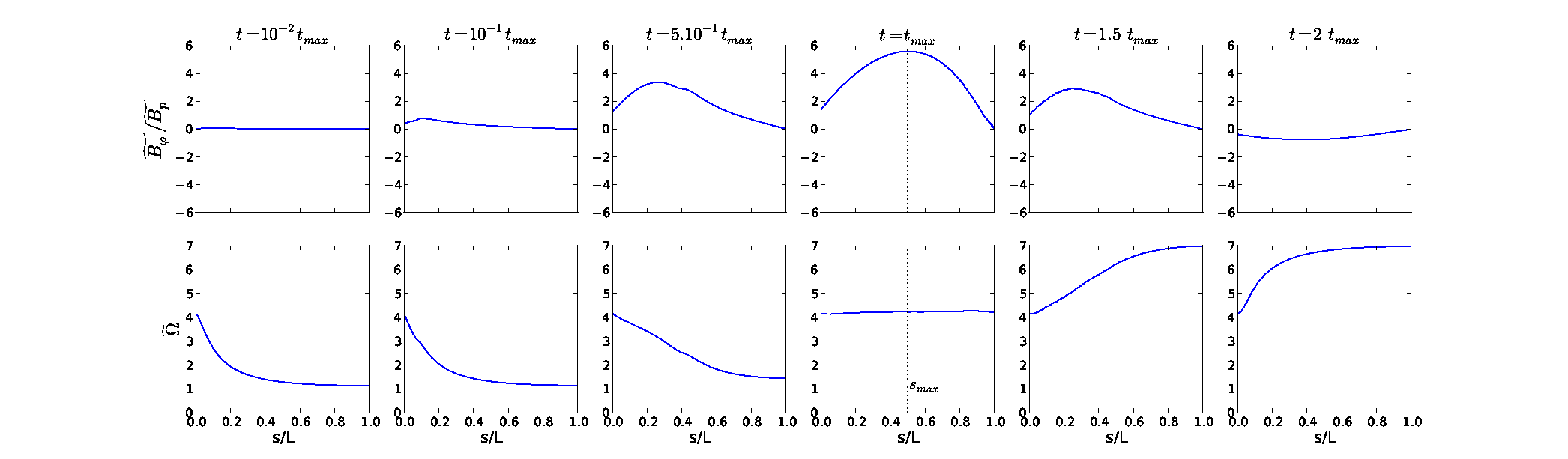}
 \caption{\small{$\widetilde{B_\varphi}/\widetilde{B_p}$ (top panels) and $\widetilde{\Omega}$ (bottom panels) as a function of $s/L$ for different times. $s$ is the curvilinear coordinate of the poloidal field line in which $max(B_\varphi/B_p)$ is reached, and $L$ the length of this field line. $s=0$ locates the inner radius and $s=L$ the equator. 
 The simulation is the same as in Fig. \ref{image2d} with the exception of $L_u=10^4$. The vertical dashed line indicates $s_{max}$ the location of $max(B_\varphi/B_p)$ on the poloidal field line.}}
 \label{bphi_lignedechamp}
\end{center}
\end{figure*}

Since non-axisymmetric instabilities are expected to occur when the magnetic configuration is dominated by the toroidal component, we now focus on the highest value of $B_\varphi/B_p$ achieved during the 
winding-up process.
In the cases we considered, this maximum is always reached before the phase-mixing episode takes place (panel $c$ for the simulation shown in Fig. \ref{image2d}).
Figure \ref{max} shows the evolution of the spatial maximal value of $B_\varphi/B_p$ for the reference case. This quantity is defined as the highest value reached in the whole domain. 
At first, we observe a nearly linear growth of the spatial maximal value of $B_\varphi/B_p$.
As can be seen in panels $a$, $b$, and $c$ of Fig. \ref{image2d}, during this phase the location of the maximum moves as an Alfv\'en wave along a poloidal field line. 
The $B_\varphi/B_p$ ratio reaches its highest value in time and space, $max(B_\varphi/B_p)$, at a time denoted $t_{max}$.
At later times in Fig. \ref{max}, the spatial maximal value of $B_\varphi/B_p$ oscillates due to reflections of Alfv\'en waves at the boundaries and rapidly decreases towards $0$ as the system evolves towards a steady state through the dissipation induced by the phase-mixing mechanism.
The temporal evolution presented in Fig. \ref{max} is typical for most of our simulations. 
We note that $(r_{max}, \theta_{max})$, the location at which $max(B_\varphi/B_p)$ is reached, is visible in panel $c$ of Fig. \ref{image2d} at mid-radius and low latitude. 
In this zone, the poloidal magnetic field lines do not reach the external radius, they are closed within the shell (see Fig. \ref{bp_omega}).

To illustrate the physical mechanisms that give rise to $max(B_\varphi/B_p)$, one can look at the evolution of $B_\varphi/B_p$ and $\Omega$ along the poloidal field line on which $max(B_\varphi/B_p)$ is reached. 
Figure \ref{bphi_lignedechamp} displays this evolution for the parameters of the reference case, except that $L_u=10^4$ instead of $10^2$.
Top panels show the growth and decrease of $B_\varphi/B_p$ between 0 and $2 \, t_{max}$, $max(B_\varphi/B_p)$ being reached at time $t_{max}$ and location $s_{max}$.
The bottom panels show that $\Omega$ behaves like a standing wave rather than a traveling wave. 
This is due to the boundary conditions used in the simulation and the spatial scale of the initial differential rotation, which is comparable to the length of the field line.
$B_\varphi$ then also behaves as a standing wave, but this is not clearly visible in the top panels where $B_\varphi/B_p$ is represented. 
Indeed, since $B_p$ is not uniform along the field line, the standing wave character of $B_\varphi$ is not obvious in $B_\varphi/B_p$.
We also note that at approximately $t=t_{max}$, the gradient of $\Omega$ projected onto the poloidal field line, $\partial\Omega/\partial{s}=\vec{e_s}\cdot\vec\nabla\Omega$ where $s$ is the curvilinear coordinate on the field line, changes sign.
According to the induction Eq. (\ref{effetomega}) in which the effects of magnetic diffusion are neglected ($L_u\gg1$), $\partial{(B_\varphi/B_p)}/\partial{t}$ changes sign together with $\partial\Omega/\partial{s}$, and we indeed observe in Fig. \ref{bphi_lignedechamp} the decrease of $B_\varphi/B_p$ after $t_{max}$.

\subsection{Summary of the parameter study}

\begin{table*}
 \centering
 \renewcommand{\arraystretch}{1.3}
 \small
 \begin{tabular}{cccccc}
 \hline
 \hline
 $L_u$ & $\widetilde{t}_{max}$ & $\widetilde{t}_{max}$ & $max(\widetilde{B_\varphi}/\widetilde{B_p})$ & $max(\widetilde{B_\varphi}/\widetilde{B_p})$ & $max(\widetilde{B_\varphi}/\widetilde{B_p})$\\
 & (simulation) & (SW model) & (simulation) & (Auri\`ere model) & (SW model) \\
 \hline
 $10^1$ & $0.76$ & $2.55$ & $0.75$& $4.86$ & $4.16$\\
 $10^2$ & $1.85$ & $2.55$ & $3.04$ & $4.86$ & $4.16$\\
 $245$ & $2.22$ & $2.55$ & $4.29$ & $4.86$ & $4.16$\\
 $10^3$ & $2.35$ & $2.55$ & $4.98$ & $4.86$ & $4.16$\\
 $10^4$ & $2.48$  & $2.55$ & $5.50$ & $4.86$ & $4.16$\\
 $10^5$ & $2.53$  & $2.55$ & $5.64$ & $4.86$ & $4.16$\\
\hline
\hline
\end{tabular}
\captionsetup{width=0.8\linewidth}
\captionof{table}{\small{Influence of $L_u$ on $\widetilde{t}_{max}$ and $max(\widetilde{B_\varphi}/\widetilde{B_p})$ obtained by the simulations and predicted by the model of Auri\`ere et al. (2007) and the model of standing waves (SW) presented in Sect. \ref{model_sw}.
The parameters of the simulations are $\rho_c/\rho_0=7\times10^3$, the initial radial differential rotation given by Eq. (\ref{ci23}) and $\Omega$ fixed at the inner radius.}}
\label{table_lu}
\end{table*}

\begin{table*}
 \centering
 \renewcommand{\arraystretch}{1.3}
 \small
 \begin{tabular}{cccccc}
 \hline
 \hline
 Density contrast & $\widetilde{t}_{max}$ & $\widetilde{t}_{max}$ & $max(\widetilde{B_\varphi}/\widetilde{B_p})$ & $max(\widetilde{B_\varphi}/\widetilde{B_p})$ & $max(\widetilde{B_\varphi}/\widetilde{B_p})$\\
 $\rho_c/\rho_0$ & (simulation) & (SW model) & (simulation) & (Auri\`ere model) & (SW model) \\
 \hline
 $1$ & $0.57$ & $0.45$ & $1.55$ & $1.00$ & $1.80$\\
 $5.6\times10^2$ & $1.15$ & $1.11$ & $2.68$ & $1.63$ & $1.94$\\
 $7\times10^3$ & $2.52$ & $2.55$ & $5.64$ & $4.86$ & $4.16$\\
 $2\times10^4$ & $3.89$ & $3.92$ & $8.55$ & $7.95$ & $6.22$\\
 $3.5\times10^4$ & $4.95$ & $5.00$ & $10.8$ & $10.4$ & $8.06$\\
 $7\times10^4$ & $6.67$ & $6.77$ & $14.5$ & $14.4$ & $11.1$\\
 $10^5$ & $7.86$ & $7.98$ & $17.0$ & $17.2$ & $12.9$\\
 $3\times10^5$ & $13.0$ & $13.3$ & $28.0$ & $29.4$ & $21.7$\\
 $10^6$ & $22.9$ & $23.5$ & $49.0$ & $53.2$ & $37.7$\\
\hline
\hline
\end{tabular}
\captionsetup{width=0.9\linewidth}
\captionof{table}{\small{Influence of the density contrast $\rho_C/\rho_0$ on $\widetilde{t}_{max}$ and $max(\widetilde{B_\varphi}/\widetilde{B_p})$.
The parameters of the simulations are $L_u=10^5$, the initial radial differential rotation given by Eq. (\ref{ci23}) and $\Omega$ fixed at the inner radius.}}
\label{table_rho}

\end{table*}

\begin{table*}

 \centering
 \renewcommand{\arraystretch}{1.3}
 \small
 \begin{tabular}{cccccc}
 \hline
 \hline
 Initial condition & $\widetilde{t}_{max}$ & $\widetilde{t}_{max}$ & $max(\widetilde{B_\varphi}/\widetilde{B_p})$ & $max(\widetilde{B_\varphi}/\widetilde{B_p})$ & $max(\widetilde{B_\varphi}/\widetilde{B_p})$  \\
 for $\Omega$ & (simulation) & (SW model) & (simulation) & (Auri\`ere model) & (SW model) \\
 \hline
 Eq. \ref{ci22} (radial profile) & $2.48$ & $2.55$ & $3.63$ & $4.58$ & $2.93$\\
 Eq. \ref{ci25} (radial profile) & $2.50$ & $2.55$ & $18.5$ & $8.85$ & $13.1$\\
 Eq. \ref{ci23} (radial profile) & $2.48$ & $2.55$ & $5.50$ & $4.86$ & $4.16$\\
 Eq. \ref{ci26} (cylindrical profile) & $0.96$ & $2.55$ & $0.31$ & $3.82$ & $0.39$\\
 Eq. \ref{ci30} (cylindrical profile) & $1.35$ & $2.55$ & $2.19$ & $6.13$ & $2.91$\\
\hline
\hline
\end{tabular}
\captionsetup{width=0.9\linewidth}
\caption{\label{table_ci}\small{Influence of the initial differential rotation on $\widetilde{t}_{max}$ and $max(\widetilde{B_\varphi}/\widetilde{B_p})$.
The parameters of the simulations are $L_u=10^4$, $\rho_c/\rho_0=7\times10^3$ and $\Omega$ fixed at the inner radius.}}
\end{table*}

\begin{table*}
 \centering
 \renewcommand{\arraystretch}{1.3}
 \small
 \begin{tabular}{ccccccc}
 \hline
 \hline
 Initial condition & Boundary condition & $\widetilde{t}_{max}$ & $\widetilde{t}_{max}$ & $max(\widetilde{B_\varphi}/\widetilde{B_p})$ & $max(\widetilde{B_\varphi}/\widetilde{B_p})$ & $max(\widetilde{B_\varphi}/\widetilde{B_p})$ \\
 for $\Omega$ & for $\Omega$ at $r=R_c$ & (simulation) & (SW model) & (simulation) & (Auri\`ere model) & (SW model) \\
 \hline
 Eq. \ref{ci23} (radial profile) & fixed value & $2.22$ & $2.55$ & $4.29$ & $4.86$ & $4.16$\\
 Eq. \ref{ci23} (radial profile) & stress-free & $1.69$ & $1.27$ & $2.97$ & $4.86$ & $3.03$\\
 Eq. \ref{ci26} (cylindrical profile) & fixed value & $0.85$ & $2.55$ & $0.26$ & $3.82$ & $0.39$\\
 Eq. \ref{ci26} (cylindrical profile) & stress-free & $0.85$ & $1.27$ & $0.26$ & $3.82$ & $0.28$\\
\hline
\hline
\end{tabular}
\captionsetup{width=0.9\linewidth}
\caption{\label{table_bc}\small{Influence of the boundary condition of $\Omega$ at the inner radius on $\widetilde{t}_{max}$ and $max(\widetilde{B_\varphi}/\widetilde{B_p})$.
The parameters of the simulations are $L_u=245$ and $\rho_c/\rho_0=7\times10^3$.}}
\end{table*}

The quantity $max(B_\varphi/B_p)$ defined in the previous section determines whether the magnetic configuration is locally dominated by the toroidal or the poloidal field, and $t_{max}$ gives the time at which $max(B_\varphi/B_p)$ is reached.
Tables \ref{table_lu}, \ref{table_rho}, \ref{table_ci}, and \ref{table_bc} summarize the values of $max(\widetilde{B_\varphi}/\widetilde{B_p})$ and $\widetilde{t}_{max}$ obtained by varying the Lundquist number $L_u$, the density contrast $\rho_c/\rho_0$, the initial differential rotation profile, and the boundary condition for $\Omega$ at $r=R_c$, respectively.
In the following, these results are discussed, starting with the effect of $L_u$.
The two last columns show the predictions of models that are presented in the next section.

\subsection{Influence of the diffusivities}
\label{diffusion}

\begin{figure}[!h]
\begin{center}
        \includegraphics[scale=0.4]{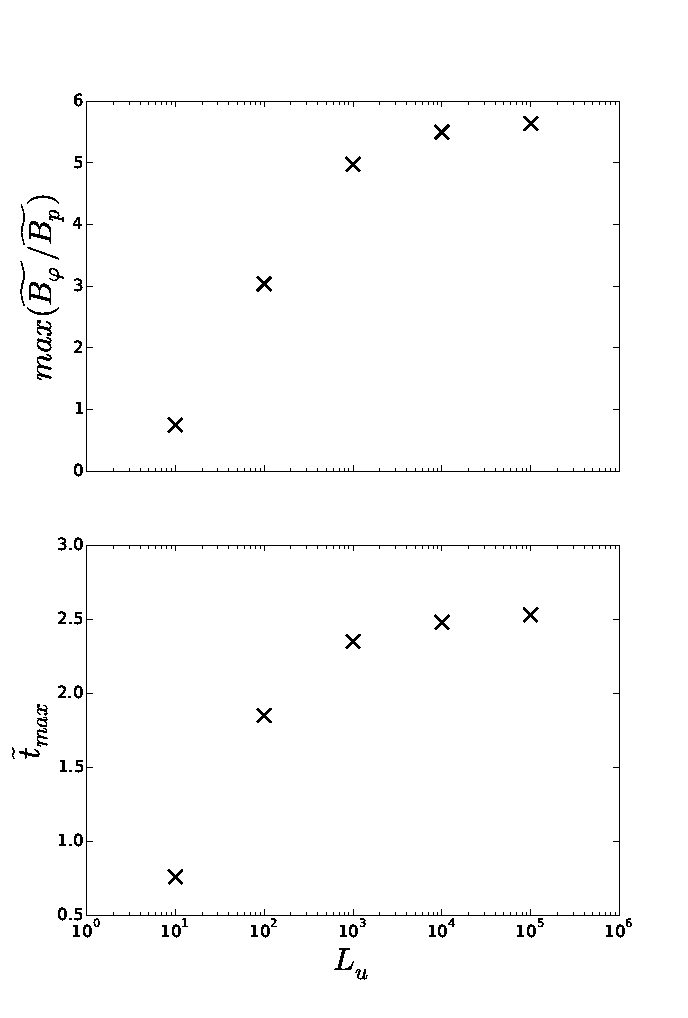}
        \caption{\small{$max(B_\varphi/B_p)$ (top) and $t_{max}$ (bottom) as a function of $L_u$ with the physical ingredients used for the reference case.}}
        \label{maxlu}
\end{center}
\end{figure}

Figure \ref{maxlu} displays $max(B_\varphi/B_p)$ and $t_{max}$ as a function of $L_u$ for the data listed in Table \ref{table_lu}.
Both $max(B_\varphi/B_p)$ and $t_{max}$ increase with $L_u$. We expect $max(B_\varphi/B_p)$ to increase when the effects of diffusion on the toroidal magnetic field decrease and thus when $L_u$ increases. To explain why $t_{max}$ increases with $L_u$, we note that  the diffusion term in the induction equation is expected to be negative (positive) near local maxima (minima) of 
$B_\varphi$. Thus, diffusion contributes to a faster change of sign of $\partial{B_\varphi}/\partial{t}$ and hence to a smaller $t_{max}$.

More interestingly, we see in this figure that an asymptotic value is reached for $L_u\gtrsim10^4$.
The same asymptotic behaviour with $L_u$ is observed for all cases considered.
The fact that the magnetic and viscous diffusions do not affect the maximum of $B_\varphi/B_p$ above a certain $L_u$ is not surprising since this maximum is reached
on a timescale of the order of an Alfv\'en time, this timescale becoming much shorter than the diffusive timescale as $L_u$ increases.
This asymptotic behaviour constitutes an interesting feature when applications to realistic stellar radiative zones are considered.

\subsection{Influence of the density profile}
\label{density_contrast}
The dependence on the density contrast was studied numerically by varying $\rho_c/\rho_0$ between $1$ and $10^6$ (see Table \ref{table_rho}). 
\begin{figure}[!h]
\begin{center}
 \includegraphics[scale=0.29,trim = 10cm 0cm 10cm 0cm]{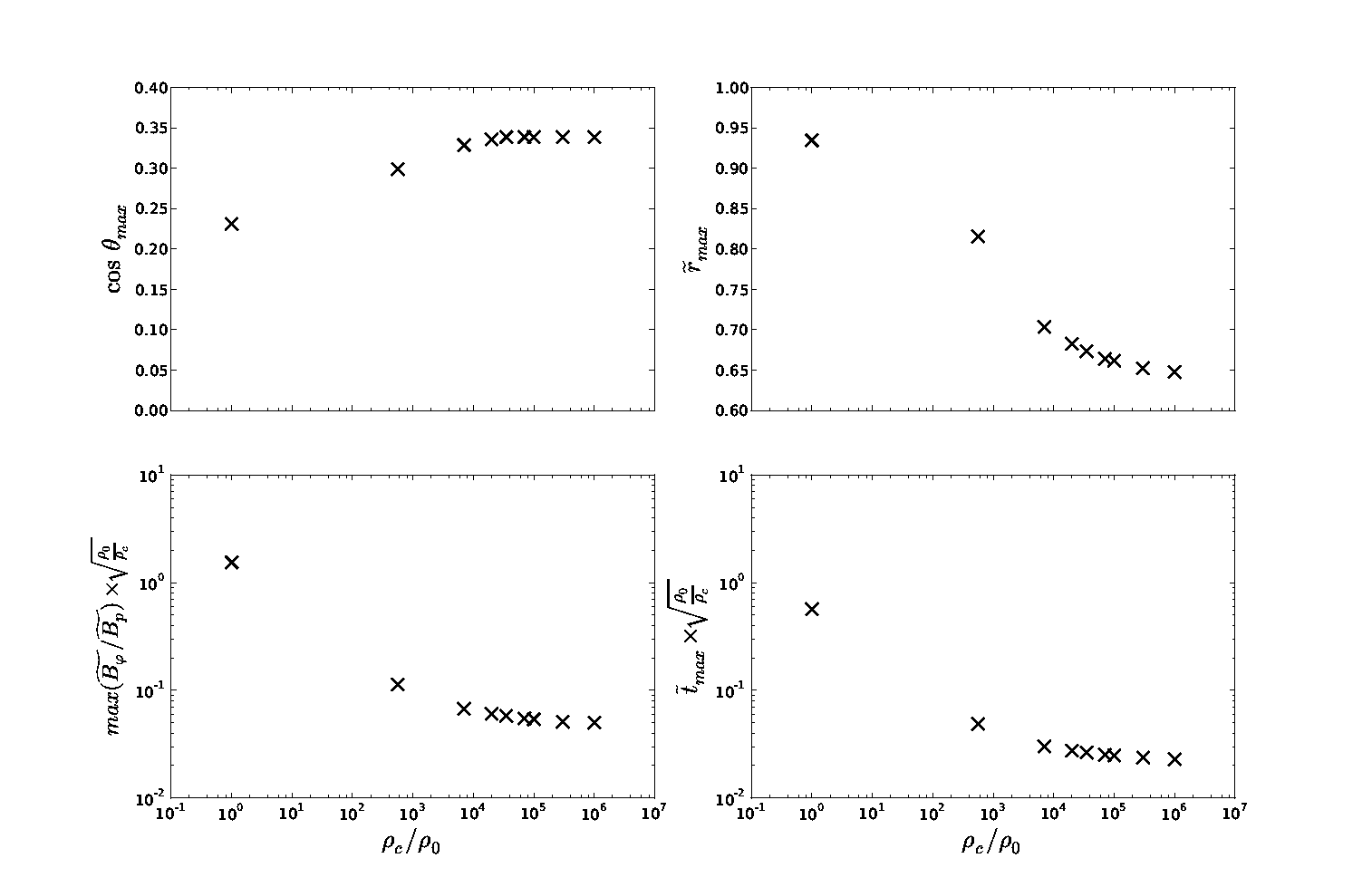}
 \caption{\small{$\cos\theta_{max}$, $\widetilde{r}_{max}$ (two top frames), $max(\widetilde{B_\varphi}/\widetilde{B_p})\times\sqrt{\rho_c/\rho_0}$ and $\widetilde{t}_{max}\times\sqrt{\rho_c/\rho_0}$ (two bottom frames) as a function of $\rho_c/\rho_0$.
 The parameters of the simulations are $L_u=10^5$, a profile of $\Omega$ defined by Eq. (\ref{ci23}) and $\Omega$ fixed at the inner radius.}}
 \label{conv_density}
\end{center}
\end{figure}
For a given configuration ($L_u=10^5$, an initial differential rotation given by Eq. \ref{ci23} and $\Omega$ fixed at the inner radius), the two top frames of Fig. \ref{conv_density} show that $\cos\theta_{max}$ and $\widetilde{r}_{max}$ respectively increases and decreases towards an asymptotic value as $\rho_c/\rho_0$ increases. 
Thus, the location of the maximum $B_\varphi/B_p$ remains unchanged above a certain $\rho_c/\rho_0$. 
As $B_\varphi$ is produced through the propagation of Aflv\'en waves along field lines, the field line where $max(B_\varphi/B_p)$ is reached is the one that goes through $\theta_{max}$ and $r_{max}$. 
This field line will remain the same above a certain $\rho_c/\rho_0$ ratio.
The same asymptotic behaviour with $\rho_c/\rho_0$ is observed for the other simulations not listed in Table \ref{table_rho}, performed for different $L_u$ and initial differential rotations.
For all the cases considered, the asymptotic value of $\widetilde{r}_{max}$ is comprised between $0.6$ and $0.8$, while $\cos\theta_{max}$ is comprised between $0.3$ and $0.4$.
Hence, $max(B_\varphi/B_p)$ is always reached in a zone located between the middle and the top of the radiative zone, close to the equator.

Table \ref{table_rho} shows that $max(\widetilde{B_\varphi}/\widetilde{B_p})$ and $\widetilde{t}_{max}$ monotonically increase with $\rho_c/\rho_0$. 
If we consider $max(\widetilde{B_\varphi}/\widetilde{B_p})\times\sqrt{\rho_0/\rho_c}$ and $\widetilde{t}_{max}\times\sqrt{\rho_0/\rho_c}$, we find that an asymptotic limit is reached for large $\rho_c/\rho_0$. 
This is shown in the two bottom frames of Fig. \ref{conv_density}.
In other words, $\frac{max(B_\varphi/B_p)}{R \, \Omega_0\sqrt{4\pi\rho_c}/B_0}\rightarrow{C1}$ and $\frac{t_{max}}{R\sqrt{4\pi\rho_c}/B_0}\rightarrow{C2}$ as $\rho_c/\rho_0$ tends towards realistic stellar values, where $C1$ and $C2$ are independent of the density contrast.
For the different simulations performed, the value of $C1$ varies between $3\times10^{-3}$ and $2\times10^{-1}$ and $C2$ between $10^{-2}$ and $3\times10^{-2}$.
The variations of these constants are mainly due to the initial differential rotation considered and, to a smaller extent, to the boundary condition adopted at the inner radius.
This asymptotic behaviour is expected if the physical mechanism that gives rise to $max(B_\varphi/B_p)$ is confined in the stellar interior since above a certain $\rho_c/\rho_0$ 
ratio the density profile is no longer modified in the internal layers. 
Indeed, $max(B_\varphi/B_p)$ is reached along a field line that remains confined in these internal layers.

\subsection{Influence of the initial and boundary conditions}
\label{icbc}

The initial differential rotation can have a strong effect on $max(B_\varphi/B_p)$, but it only slightly modifies $t_{max}$. 
Indeed, as observed in Table \ref{table_ci}, the value of $max(B_\varphi/B_p)$ varies from $18.5 \, t_{Ap}/t_\Omega$ for the $1/r^2$ profile (Eq. \ref{ci25}), which is the strongest shear we can use that is hydrodynamically stable, to $0.31t_{Ap}/t_\Omega$ for the cylindrical profile given by Eq. (\ref{ci26}). 
Generally, the stronger the shear, the larger $max(B_\varphi/B_p)$. 
But we also find that the cylindrical differential rotation induces less $B_\varphi$ than the radial one because the isocontours of $\Omega$ are more aligned with the poloidal magnetic field lines in the cylindrical case, causing the $\Omega$-effect to be less efficient.
The value of $t_{max}$ is approximately independent of the radial differential rotation profile. This indicates that the time to reverse the shear does not seem to depend on its intensity.
However, $t_{max}$ varies from $0.96 \, t_{Ap}$ to $1.35 \, t_{Ap}$ for the two cylindrical profiles considered.

Our variables of interest $max(B_\varphi/B_p)$ and $t_{max}$ are less influenced by the boundary conditions on $\Omega$ than by all the other parameters considered in this study. 
As illustrated in Table \ref{table_bc}, for radial differential rotation profiles, $max(B_\varphi/B_p)$ and $t_{max}$ are always slightly larger for a fixed value than for a stress-free boundary condition, $1.01$ to $1.66$ and $1.03$ to $1.37$ times higher, respectively, depending on the cases.  
For the cylindrical profiles, $max(B_\varphi/B_p)$ and $t_{max}$ reach the same values, independently of the boundary conditions. 

The boundary conditions other than the one on $\Omega$ at $r=R_c$ only have a limited impact on $max(B_\varphi/B_p)$ and $t_{max}$.
We thus do not show their influence here. If longer timescales were considered (and final steady states), they would play a significant role, however.

\section{Simple models}
\label{models}
In this section two models are presented that provide an estimate of $max(B_\varphi/B_p)$ and $t_{max}$. Their predictions are compared with the results of the simulations.
The first model is a local model that uses the same phenomenological assumptions as \cite{auriere2007}. 
The second model consists of estimating the poloidal field line on which $max(B_\varphi/B_p)$ is reached and then of solving a 1D Alfv\'en wave equation along this field line.

\subsection{Local model of Auri\`ere et al. (2007)}
\label{modele}

Starting from the ideal induction equation for the toroidal component
\begin{equation}
\label{ind_model}
 \frac{\partial{B_\varphi}}{\partial{t}}=r\sin\theta \, \vec{B_p}\cdot\vec{\nabla}\Omega\ \ \ ,
\end{equation}
\cite{auriere2007} assumed that the toroidal field back-reacts on the differential rotation after a time $\tau_{max}=\ell/v_{Ap}$, corresponding to the period of an Alfv\'en wave of wavelength $\ell=|\vec{\nabla}\Omega|/\Omega$, the length scale of the
initial differential rotation. Then, by integrating Eq. (\ref{ind_model}) from $t=0$ to $t=\tau_{max}$ and assuming that the differential rotation is constant up to that time, we obtain
\begin{equation}
 \label{model_eq}
 \frac{B_\varphi}{B_p}(r,\theta,t=\tau_{max})=\frac{r\sin\theta \, \Omega(r,\theta,t=0)}{v_{Ap}(r,\theta)}\ \ \ .
\end{equation}
Taking the spatial maximum of $B_\varphi/B_p$ in Eq. (\ref{model_eq}), $max(B_\varphi/B_p), r_{max}, \theta_{max}$ and then $\tau_{max}$ can be determined.

Of the assumptions of this model, neglecting magnetic diffusion appears justified since the numerical results have shown that $max(B_\varphi/B_p)$ becomes independent of $L_u$ when $L_u\gtrsim10^4-10^5$.
Relating $\tau_{max}$ to the length scale of the initial differential rotation $\ell$ leads to a simple expression, but then $B_\varphi/B_p$ taken at $\tau_{max}$ does not depend on the 
initial gradient of $\Omega$ and in particular on its projection onto poloidal field lines. 
For example, if Ferraro's law is verified initially, $B_\varphi$ should remain equal to zero, but the model incorrectly predicts that induction takes place.
Another problem is that with a local model, the effects of the boundary conditions are not taken into account.

\subsection{Model of standing Alfv\'en waves on a poloidal magnetic field line}
\label{model_sw}

Since the initial perturbations of $\Omega$ and $B_\varphi$ propagate as Alfv\'en waves along $\vec{B_p}$, a more accurate model would be to solve 
a one-dimensional wave equation along the poloidal field line where  $max(B_\varphi/B_p)$ is reached. 
As we expect this field line to be the one for which the $\Omega$-effect lasts the longest, we assume that $max(B_\varphi/B_p)$ is reached on the field line that maximizes the Alfv\'en propagating time over
the line length $L$ (taken from the inner radius to the equator or the surface) $\tau(L)=\int_{0}^{L}ds/v_{Ap}$.
The field line obtained using this condition for a case with $\rho_c/\rho_0=7\times10^3$ is indicated in red in Fig. (\ref{bp_omega}).

Neglecting the magnetic and viscous diffusions and assuming that the scale of variation of $r\sin\theta$ and $B_p$ along $\vec{B_p}$ is larger than the scale of $\partial\Omega/\partial{s}$ \citep{MW87},
the 1D wave equation to be solved on this field line is
\begin{equation}
 \label{wave_equation}
 \frac{\partial^2\Omega(s,t)}{\partial{s}^2}-\frac{1}{v_{Ap}^2(s)}\frac{\partial^2\Omega(s,t)}{\partial{t}^2}=0\ \ \ .
\end{equation}
%taking into account the boundary and the initial conditions.

As already noticed, $\Omega$ behaves approximately as a standing wave for the initial differential rotation considered here (see Fig. \ref{bphi_lignedechamp}). 
We thus search for normal-mode solutions $ f(s)\exp(i\omega{t})$ of the wave equation. 
In addition, we consider the fundamental mode as an approximate solution of the problem because the initial perturbation is dominated by its large-scale component. 
The equation for $f(s)$ is
\begin{equation}
 \label{f_equation}
 \frac{d^{2}f(s)}{ds^2}+\left(\frac{\omega}{v_{Ap}(s)}\right)^{2}f(s)=0\ \ \ .
\end{equation}

A simple and explicit solution is obtained assuming that $v_{Ap}$ is a constant equal to the mean Alfv\'en velocity $\overline{v_{Ap}}$ along the poloidal field line. 
We then obtain $B_\varphi/B_p(s,t=t_{max})$ by integrating the induction equation up to $t_{max}$
\begin{equation}
 \label{solution_fixed_const}
 \frac{B_\varphi}{B_p}(s,t=t_{max})=\Delta\Omega \, \frac{g(s)}{\overline{v_{Ap}}} \, \cos\left(\frac{\pi}{2}\frac{s}{L}\right)\ \ \ ,
 \end{equation}
for a fixed boundary condition at $s=0$ and a stress-free boundary condition at $s=L$, with $t_{max}=\frac{L}{\overline{v_{Ap}}}$ and where $\Delta\Omega=\Omega(s=0,t=0)-\Omega(s=L,t=0)$ and $g(s)=r(s)\sin\theta(s)$.
\begin{equation}
 \label{solution_stressfree_const}
 \frac{B_\varphi}{B_p}(s,t=t_{max})=\frac{\Delta\Omega}{2} \, \frac{g(s)}{\overline{v_{Ap}}} \, \sin\left(\pi\frac{s}{L}\right)\ \ \ ,
\end{equation}
for a stress-free boundary condition at $s=0$ and $s=L$ and with $t_{max}=\frac{L}{2\overline{v_{Ap}}}$.

A more accurate solution can be obtained using the Wentzel-Kramers-Brillouin (WKB) approximation
\begin{equation}
 \label{solution_fixed}
 \frac{B_\varphi}{B_p}(s,t=t_{max})=\Delta\Omega \, \frac{g(s)}{\sqrt{v_{Ap}(s) \, v_{Ap}(L)}} \, \cos\left(\frac{\pi}{2}\frac{\tau(s)}{\tau(L)}\right)\ \ \ ,
\end{equation}
for a fixed boundary condition at $s=0$ and a stress-free boundary condition at $s=L$, with $t_{max}=\tau(L)$ and where $\tau(s)=\int_{0}^{s}\frac{ds'}{v_{Ap}}$.
\begin{equation}
 \label{solution_stressfree}
 \frac{B_\varphi}{B_p}(s,t=t_{max})=\frac{\Delta\Omega}{2} \, \frac{g(s)}{\sqrt{v_{Ap}(s) \, v_{Ap}(L)}} \, \sin\left(\pi\frac{\tau(s)}{\tau(L)}\right)\ \ \ ,
\end{equation} 
for a stress-free boundary condition at $s=0$ and $s=L$ and with $t_{max}=\frac{\tau(L)}{2}$.

We now compare this WKB solution and the model of Auri\`ere et al. (2007), hereafter called SW model for "standing wave" and Auri\`ere model, with the results of the numerical simulations.

\subsection{Comparison of the models with the numerical simulations}

\begin{figure*}
\begin{center}
 \includegraphics[scale=0.38,trim = 11cm 0cm 10cm 0cm]{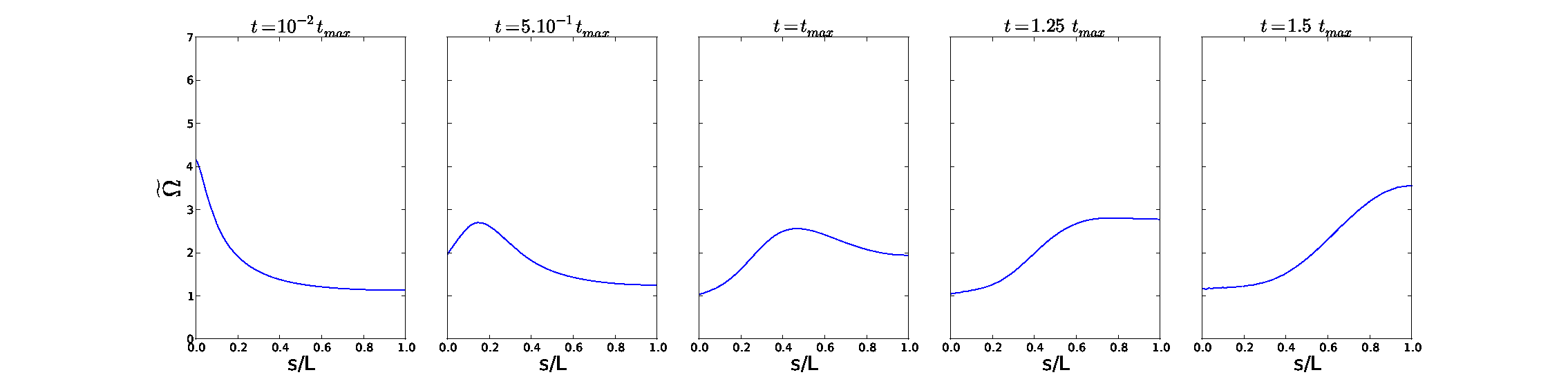}
 \caption{\small{$\widetilde{\Omega}$ as a function of $s/L$ for different times. $s$ is the curvilinear coordinate of the poloidal field line on which $max(B_\varphi/B_p)$ is reached, and $L$ the length of this field line. $s=0$ locates the inner radius and $s=L$ the equator. 
 The parameters of the simulation are $L_u=245$, $\rho_c/\rho_0=7\times10^3$, the differential rotation profile given by Eq. (\ref{ci23}) and a stress-free boundary condition.}}
 \label{omega_stressfree}
\end{center}
\end{figure*}

The values of $max(B_\varphi/B_p)$ and $t_{max}$ given by the models are compared with the results of the simulations in Tables \ref{table_lu}, \ref{table_rho}, \ref{table_ci} and \ref{table_bc}.

First, we observe that the Auri\`ere model and the SW model reproduce the asymptotic behaviour of $max(B_\varphi/B_p)$ when $\rho_c/\rho_0$ increases. 
Indeed, Table \ref{table_rho} indicates that $max(B_\varphi/B_p)$ is given with a mean relative error of $13\%$ by the  Auri\`ere model against $24\%$ for the SW model.
In this particular case, the Auri\`ere model agrees better with the simulations than the SW model. 
However, in the other simulations that are not listed in this paper it is generally the opposite trend.
The SW model provides very accurate predictions of $t_{max}$ with a mean relative error of only $4\%$, while the Auri\`ere model predicts values of $t_{max}$ several orders of magnitude higher than the simulations (we have thus chosen not to list them in the tables).

As expected, the effect of the initial differential rotation is much better reproduced by the SW model than by the Auri\`ere model, with a mean relative error on $max(B_\varphi/B_p)$ of $26\%$ and $280\%$, respectively (see Table \ref{table_ci}).
While the Auri\`ere model does not take into account the effects of the boundary conditions, the SW model estimates $max(B_\varphi/B_p)$ with a mean relative error of only $16\%$ for the different boundary conditions considered (see Table \ref{table_bc}). 
Finally, Fig. \ref{omega_stressfree}, for the stress-free BC case, and Fig. \ref{bphi_lignedechamp}, for the fixed BC, indeed show that the evolution of the perturbation is not far from the standing waves prescribed by the SW model.

We conclude that the SW model provides a useful approximation for the maximum ratio $B_\varphi/B_p$ in a differentially rotating star with a dipolar field, 
provided that the length scale of the initial differential rotation is not too small compared to the stellar radius.

\section{Towards the instability of the magnetic field}
\label{instability}

\begin{figure}[!h]
\begin{center}
 \includegraphics[scale=0.035]{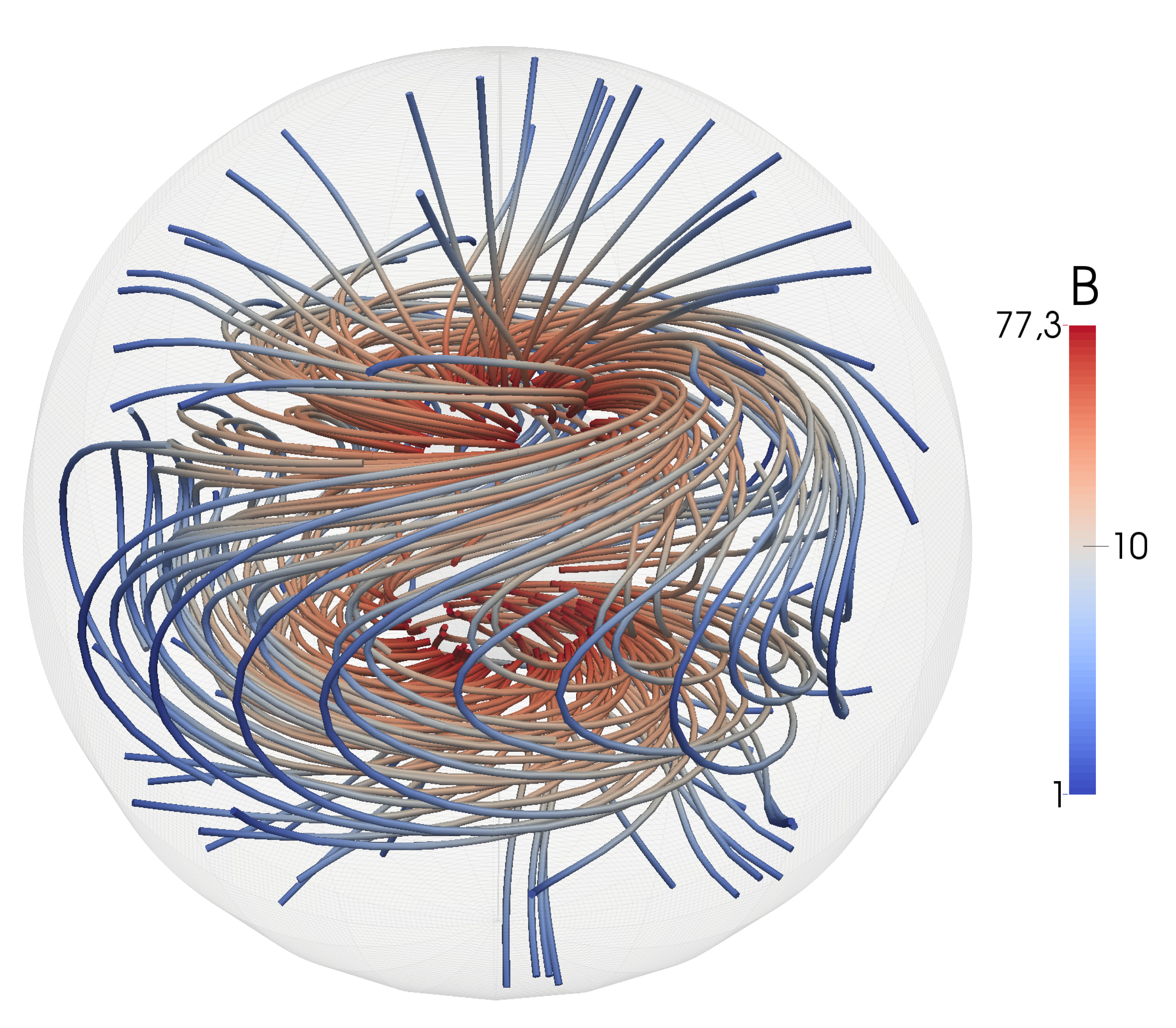}
 \caption{\small{3D view of the magnetic configuration obtained with our 2D simulations at $t_{max}$. 
 It is computed for $L_u=10^2$, a density contrast $\rho_c/\rho_0=7\times10^3$, the radial differential rotation profile defined by Eq. (\ref{ci23}) and $\Omega$ fixed at the inner radius.}}
 \label{3d}
\end{center}
\end{figure}

We now turn to the discussion of possible instabilities of the magnetic configurations found in our simulations. 
Figure \ref{3d} shows the configuration obtained at time $t_{max}$ in a typical simulation. A 3D rendering of the field lines is shown, coloured by the total magnetic intensity. 
We clearly see in this figure that we have a complex structure with mixed poloidal and toroidal components, even if the toroidal field dominates at mid-latitudes. 
Moreover, differential rotation is still present in the spherical domain at this time of the simulation and may strongly influence the stability conditions of this complex magnetic field.

One way to study the stability of such a complex magnetic configuration without approximation is to perform 3D numerical simulations where the full set of MHD equations is solved. 
This approach has been followed in a companion paper \citep{jouve2015}, although the magnetic configurations studied in 3D have been limited to cases of uniform density, 
cylindrical differential rotation, and low $L_u$ values.
\cite{jouve2015} found that a magnetic instability is triggered and destroys the large-scale magnetic field if the ratio of the poloidal Alfvén time $t_{Ap}$ to the rotation timescale $t_\Omega$ is sufficiently high.
In unstable cases, an enhanced transport of angular momentum due to the turbulence induced by the instability is found, confirming the results of \cite{rudiger2015} in cylindrical geometry.
This fully 3D study also served to test the capacity of approximate local criteria to determine the stability of these complex configurations. 
In particular, it was found that the dispersion relation of \cite{ogilvie2007} could predict the nature of instability and provide reasonable estimates 
of its growth rates. Here, we used these local criteria to predict the stability of the configuration obtained through our axisymmetric simulations. 
As compared to \cite{jouve2015}, we therefore study the stability of a wider range of configurations, including the effects of the density stratification, of a radial differential rotation profile, and of stable stratification.

\subsection{Local stability analysis under simplifying assumptions}

From numerous previous studies, we know that a magnetic field is more likely to be unstable if it is dominated by one of its components, namely when it is either purely poloidal \citep{MT73,FR77} or purely toroidal \citep{tayler73,PT85}. 
In the previous sections, we have established in which conditions the initial differential rotation produces a configuration dominated by the toroidal field. 
For the stability analysis, we only consider the situations where the toroidal magnetic field is so dominant that the poloidal component can be neglected. 
We return to this assumption in the next sub-section.

A major difficulty in performing a linear stability analysis of our magnetic configurations is that the axisymmetric background field evolves in time. 
Therefore, an important quantity to consider here is the ratio between the growth time of possible instabilities and the evolution timescale of the background magnetic field. 
From now on, we assume that this ratio is low and thus that the background state can be assumed to be steady. 
This assumption is discussed in the next sub-section.

The local magnitude of the differential rotation is measured by $q \Omega$, where $q=\partial \ln \Omega/\partial \ln \varpi$ is the shear parameter and $\varpi$ is the cylindrical radius. 
For the different profiles considered (see Eqs. \ref{ci22} to \ref{ci30}), the shear parameter is always of order unity, meaning that the differential rotation is of the order of the local rotation rate. 

\bigskip

Under these assumptions, we may proceed to a local linear stability analysis, using the dispersion relation derived by \cite{ogilvie2007} for the case of a purely toroidal field under the influence of a differential rotation, both possessing arbitrary profiles. 
Two instabilities may be expected in this situation: the Tayler instability (TI), which derives from free magnetic energy, and the magneto-rotational instability (MRI), whose source of energy is the free kinetic energy of the differential rotation. 
The steady and axisymmetric basic state consists of a differential rotation profile $\Omega(\varpi,z)$ and a purely toroidal magnetic field $B_\varphi(\varpi,z)$, where $z$ is the coordinate along the rotation axis. 
If we consider a perturbation such that the displacement is in the direction $\vec{e}$, we obtain the following form of the dispersion relation:
\begin{align}
\begin{split}
&\left [ \omega^2-\frac{m^2 v_{A\varphi}^2}{\varpi^2} -(\vec{e}\cdot\vec{e_r})^2 N^2-2\varpi \, \Omega \, {\vec{e}}\cdot{\vec{\nabla}}\Omega+\frac{2 \, v_{A\varphi}}{\sqrt{4\pi\rho}} \, {\vec{e}} \cdot {\vec{\nabla}}\left( \frac{B_\varphi}{\varpi} \right)\right ]\\
&\times\left[ \omega^2-\frac{m^2v_{A\varphi}^2}{\varpi^2}  \right] = \left[ 2  \, \Omega \, \omega + \frac{2m \, v_{A\varphi}^2}{\varpi^2}  \right]^2 \ \ \ ,
\end{split}
\end{align}
where $\omega$ is the frequency of the perturbation, $m$ is the azimuthal wavenumber, $v_{A\varphi}=B_\varphi/\sqrt{4\pi\rho}$ is the toroidal Alfv\'en velocity, and $N$ is the Brunt-V\"ais\"al\"a frequency taken as uniform in the whole domain.
As noted previously, the buoyancy force does not affect the evolution of the axisymmetric magnetic field since the meridional flows are neglected. 
It influences the stability of the magnetic configurations through the value of the $N/\Omega$ ratio, however. 
As noted by \cite{jouve2015}, another important quantity of the magnetic stability is the ratio of the rotation rate $\Omega$ to the toroidal Alfv\'en frequency $\omega_{A{\varphi}}=v_{A\varphi}/r$. 

\begin{figure}[!h]
\begin{center}
 \includegraphics[scale=0.1]{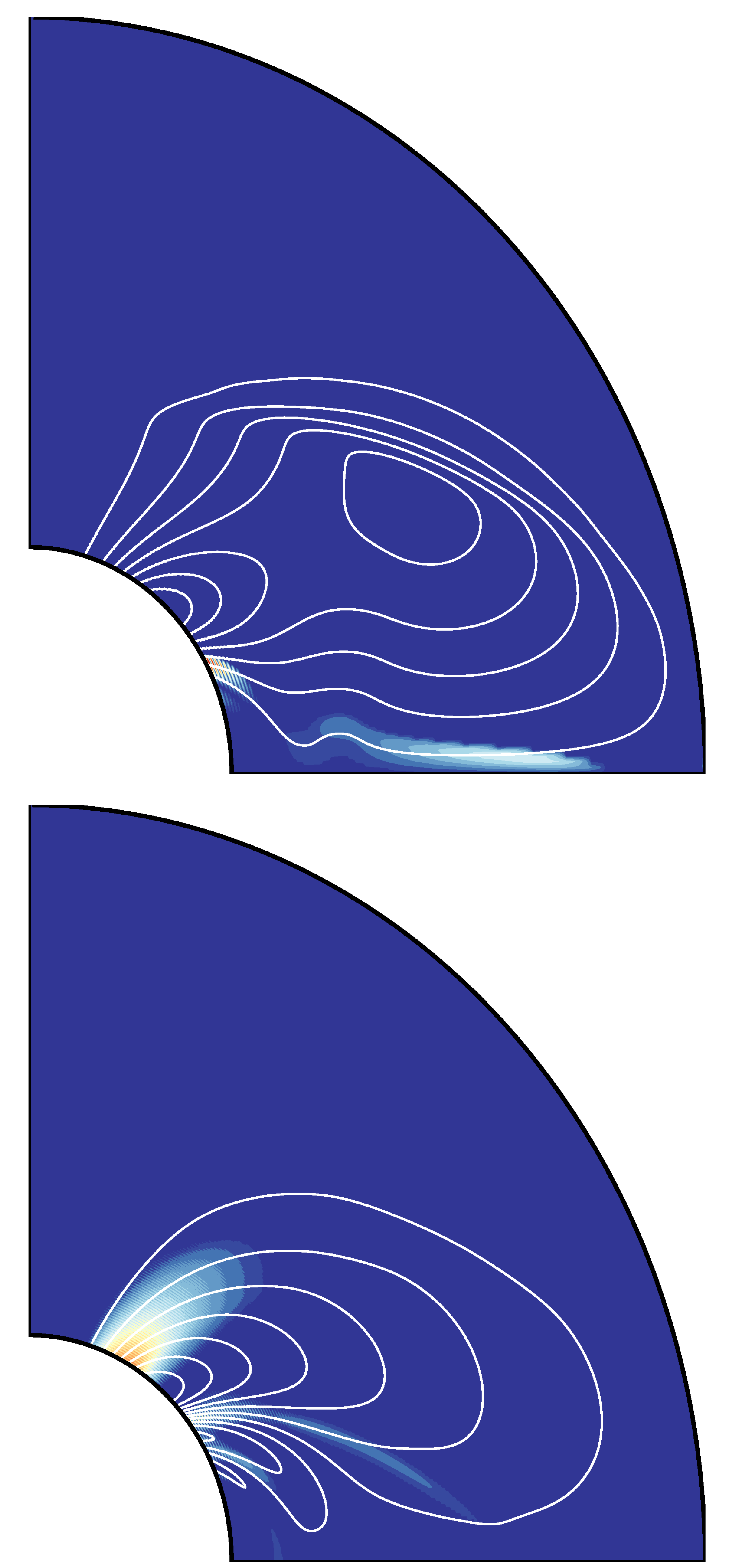}
 \caption{\small{Contours of the toroidal magnetic field (white lines) and of the growth rate of the instability (coloured contours) for the most unstable perturbation ($m=7$ for the top frame and $m=1$ for the bottom frame) which is in the direction $\vec{e}=\vec{e_\varpi}$. 
 The parameters of the simulations are $N=0$ and, respectively for the top and the bottom frame $\Omega/\omega_{A{\varphi}}\approx 10$ and $\Omega/\omega_{A{\varphi}}\approx 1$ at the maximum of the toroidal field.}}
 \label{fig_tivsmri}
\end{center}
\end{figure}

For a given configuration $B_\varphi(\varpi,z,t_{max})$ and $\Omega(\varpi,z,t_{max})$, corresponding to a $\Omega(\varpi,z,t_{max})/\omega_{A\varphi}(\varpi,z,t_{max})$ ratio, obtained with the 2D numerical simulations and a value of $N/\Omega$, the dispersion relation provides the growth rate $\sigma=\Im(\omega(\varpi,z))$ of a local perturbation characterized by $m$ and $\vec{e}$.
Two cases are illustrated in Fig. \ref{fig_tivsmri}. In this figure, only the northern hemisphere is shown since the results are symmetric with respect to the equator. 
The contours of the toroidal magnetic field are drawn (white lines) together with the contours of the growth rate of the instability (coloured contours) for the most unstable perturbation, which is in the direction $\vec{e}=\vec{e_\varpi}$. 
The only difference in the initial setup of these two simulations is the differential rotation profile (Eq.\ref{ci26} was used for the case in the top panel and Eq.\ref{ci23} for the bottom panel). 
As a consequence of a different winding-up between these two cases, the ratio $\Omega/\omega_{A{\varphi}}$ reaches very different values. 
As shown in \cite{jouve2015}, the nature of the instability that is likely to be triggered in these magnetic configurations depends on this $\Omega/\omega_{A{\varphi}}$ frequency ratio. 
For the case in the top panel $\Omega/\omega_{A{\varphi}}\approx 10$ at the maximum of toroidal field, the differential rotation thus plays a dominant role, and the most vigorous instability is the MRI, the most unstable mode having an azimuthal wave number $m=7$. 
In contrast, for the other case where $\Omega/\omega_{A{\varphi}}\approx 1$ at the maximum of toroidal field, the current-driven Tayler instability is triggered and favours the $m=1$ mode. The location of the instability is also quite different in both cases. 
In the MRI case, the most unstable region is concentrated at rather low latitudes and very extended in radius. 
In the TI case, however, the unstable zones are mainly located on strong gradients of the field close to the bottom of the domain at latitudes $30^o$ and $60^o$. 
In all cases studied, the value of $\Omega/\omega_{A{\varphi}}$ is found to vary between about $1$ and $10$ depending on the initial and boundary conditions. 
We thus expect our axisymmetric configurations to be subject to an instability, the nature of which only depends on this $\Omega/\omega_{A{\varphi}}$ ratio. 
Both the MRI and the TI are found to be likely to exist in our situations. 
The location of the instability, the growth rates as well as the typical lengthscale of the most unstable mode will then differ and the consequences on the observed surface field may be very different between the TI and the MRI cases.

We showed in Sect. \ref{density_contrast} that when the density contrast is increased between the top and bottom of our domain, both the highest value of $B_\varphi/B_p$ and its location tend to an asymptotic value. 
For the stability conditions, we find that for the case where the $m=1$ Tayler instability is favoured (bottom panel of Fig. \ref{fig_tivsmri}), increasing the density contrast does not significantly vary the nature, location, and growth rate of the instability. 
Indeed, as $\rho_c/\rho_0$ is increased from $7\times10^3$ (case shown in the figure) to $2\times10^4$ and to $10^6$, the whole magnetic configuration is left mostly unchanged and the value of $\Omega/\omega_{A{\varphi}}$ increases from $1$ to $1.25$ and to $1.54$. 
In these three cases, the $m=1$ mode is found to be the most unstable to the TI and is always located at the base of the domain in two distinct regions in latitude (i.e. at $30^o$ and $60^o$ where the gradients are the strongest). 
The highest growth rate is always around $\sigma\approx 7-9 \, \Omega_{A{\varphi}}$, where $\Omega_{A\varphi}=B_{\varphi_m}/R\sqrt{4\pi\rho_c}$ and $B_{\varphi_m}$ is the maximal toroidal field.
This is an interesting feature of this work since it implies that our results can be extrapolated to realistic values of the density contrast not only for the location and value of $max(B_\varphi/B_p)$, but also for the instabilities that are likely to be triggered in the stellar interior.

The effect of the stratification has also been explored by varying the parameter $N/\Omega_0$, where $\Omega_0$ is the rotation rate at the surface.
For the case where the Tayler instability is favoured, increasing $N/\Omega_0$ from $0$ (case shown in the bottom panel of Fig. \ref{fig_tivsmri}) to $8$ and to $16$ decreases the maximum growth rate of the instability from $9.2 \, \Omega_{A{\varphi}}$ to $8.2 \, \Omega_{A{\varphi}}$ and to $5.2 \, \Omega_{A{\varphi}}$.
The location of the instability remains unchanged.
For the case where the MRI is favoured, the highest growth rate strongly decreases, the most favoured wave number $m$ decreases when $N/\Omega_0$ increases, and the instability totally disappears for $N/\Omega_0 \ge 2$.
This agrees with \cite{spruit99}, who stated that the MRI is more affected by the stable stratification than the TI. 
It also agrees with \cite{masada2006}, who showed that the most unstable azimuthal wave number $m$ decreases when $N/\Omega_0$ increases.
Nevertheless, with a realistic stable stratification ($N/\Omega_0\simeq10$), the most unstable perturbations are no longer in the direction $\vec{e}=\vec{e_\varpi}$, but rather in the direction $\vec{e}=\vec{e_\theta}$.
Indeed, such horizontal perturbations are not affected by the stratification, and magnetic instabilities are still present in these cases.
For the same case as the one shown in the bottom panel of Fig. \ref{fig_tivsmri}, the $m=1$ TI still possesses a significant growth rate $\sigma\simeq9 \, \Omega_{A\varphi}$ in the direction $\vec{e}=\vec{e_\theta}$.
The MRI can still be triggered for $\vec{e}=\vec{e_\theta}$, but with a growth rate $\sigma\simeq0.4 \, \Omega_{A\varphi}$ much lower than in the unstratified cases, where $\sigma\simeq2.6 \, \Omega_{A\varphi}$ (case shown in the top panel of Fig. \ref{fig_tivsmri}).
We thus still expect non-axisymmetric instabilities to develop in the stably stratified cases, the main difference being that when $\Omega/\omega_{A{\varphi}}>1$ (MRI regime), the growth rate is significantly reduced.

\subsection{Discussion of the stability condition}

One of the simplifying assumptions in the previous section was to consider that the instability would grow on a steady axisymmetric background state. 
However, both the background magnetic field and the differential rotation always oscillate on the short timescales considered in our study. 

From Fig. \ref{max}, this oscillation frequency can be estimated to be $1/2 \,t_{max}$.
To freely develop, an instability with a growth rate $\sigma$ should therefore fulfil the condition
\begin{equation}
 \label{growth}
 \sigma>\frac{1}{2 \, t_{max}} \ \ \ .
\end{equation}

As discussed in the previous sub-section, if $\Omega/\omega_{A\varphi}>1$, the MRI is favoured. Its maximum growth rate is related to the shear parameter and the rotation rate in the following way \citep[see for example][]{BH91}:
\begin{equation}
\sigma_{MRI}=q \, \Omega/2 \ \ \ .
\end{equation}
If $\Omega/\omega_{A\varphi}\lesssim1$, the TI is favoured and the growth rate reads
\begin{equation}
 \sigma_{TI}=\omega_{A\varphi}\ \ \ .
\end{equation}

From the standing Alfv\'en wave model presented in Sect. \ref{model_sw}, the toroidal Alfv\'en frequency $\omega_{A\varphi}$ and $t_{max}$ are given by Eqs. (\ref{solution_fixed}) and (\ref{solution_stressfree}), or by Eqs. (\ref{solution_fixed_const}) and (\ref{solution_stressfree_const}) in a simplified case.
The quantity $\Delta\Omega$, which appears in these equations, can be expressed in terms of the rotation rate $\Omega$ and the shear parameter $q$: $\Delta\Omega\approx \varpi \, \partial \Omega/\partial \varpi= q \, \Omega$.
The condition for the development of a non-axisymmetric instability, Eq. (\ref{growth}), can then be rewritten as
\begin{equation}
 \label{growth_model}
 \Delta\Omega > \frac{\overline{v_{Ap}}}{L} \ \ \ ,
\end{equation}
which is valid for both the MRI and the TI cases.

In practice, effects ignored in this estimation will make this condition more restrictive. 
First, as shown in the previous sub-section, the stable stratification is expected to significantly decrease the instability growth rate in the MRI case. 
Second, to derive (\ref{growth_model}), we assumed that the toroidal component dominates the poloidal component. 
This assumption does not impose an additional constraint since, according to the SW model, condition (\ref{growth_model}) also ensures that the toroidal component dominates.
Nevertheless, even in the presence of a small poloidal component, a more conservative instability condition might be necessary \citep[see for the TI case][]{linton96,braithwaite2009}.
Finally, the 3D simulations by \cite{jouve2015} indicate that a condition $\sigma>10 \frac{1}{2 \, t_{max}}$ is more appropriate than  $\sigma>\frac{1}{2 \, t_{max}}$ to account for the effect of the oscillating background state.     
Indeed, in the regime $\frac{1}{2 \, t_{max}} <\sigma<10 \frac{1}{2 \, t_{max}}$ perturbations do grow exponentially, but are killed by the background field reversal before they reach the level of energy of the axisymmetric field.
Taking these three effects into account, the condition for the  non-axisymmetric instabilities therefore is rather $\Delta\Omega >> \frac{\overline{v_{Ap}}}{L}$. 
In the particular case of the 3D simulations by \cite{jouve2015}, the condition is $\Delta\Omega > 100 \frac{\overline{v_{Ap}}}{L}$.

\subsection{Discussion of the critical surface field}
\label{sect_threshold}

When written in terms of quantities observables at the stellar surface, the condition for instability reads

\begin{equation}
 \label{condition_b0}
\frac{B_0}{R \Omega_0\sqrt{4\pi\rho_0}} < C \frac{\Delta\Omega}{\Omega_0} \frac{v_{Ap}(R)}{\overline{v_{Ap}}}	\ \ \ ,
\end{equation}
where, as we just discussed, the factor $C$ takes into account effects on the instability ignored in the simple condition (\ref{growth_model}).
In the case studied by 3D numerical simulations (where $\rho$ is uniform, $\Delta\Omega/\Omega_0 \sim 1$ and the initial field is dipolar), it is found to be $\sim 10^{-2}$.
This expression defines a surface critical field, $B_{crit}$, below which non-axisymmetric instabilities can be triggered by differential rotation.
This field is proportional to $R \Omega_0\sqrt{4\pi\rho_0}$ and depends on internal properties, such as differential rotation and field geometry, through the ratio $\Delta\Omega/\Omega_0$ and $v_{Ap}(R)/\overline{v_{Ap}}$. 
The actual value of $\Delta\Omega/\Omega_0$ will depend on the mechanism that enforces differential rotation in the star and might be much lower than the $\Delta\Omega/\Omega_0\sim 10^{-1}-1$ values we considered here as initial conditions.

Following \cite{auriere2007}, the magnetic dichotomy of intermediate-mass stars would be due to the development of non-axisymmetric instabilities separating stable strong field configurations, the Ap/Bp stars, from unstable weaker field configurations whose surface average field becomes very weak after the destabilization. 
Accordingly, the lower limit of Ap/Bp magnetic fields would be given by a stability condition such as (\ref{condition_b0}) (the present stability condition is global and thus differs from the local condition found by \cite{auriere2007}).
As we know that the observed lower bound of Ap/Bp dipolar fields is $\simeq 300  \, {\rm G}$ and that $R \, \Omega_0\sqrt{4\pi\rho_0} \simeq 300 \, {\rm G}$ for a typical Ap star ($R=3 \, R_\odot$, $P=5 \, {\rm days}$, $\log g = 4  \, {\rm dex}$ and $T_{\rm eff} = 10^4  \, {\rm K}$), 
the condition for instability (\ref{condition_b0}) would match observations if $ C \frac{\Delta\Omega}{\Omega_0} \frac{v_{Ap}(R)}{\overline{v_{Ap}}} \gtrsim 1$. 
A density profile calculated for such a star thanks to the stellar evolutionary code MESA yields $\frac{v_{Ap}(R)}{\overline{v_{Ap}}}\simeq2\times10^4$, which implies that $\Delta\Omega/\Omega_0\gtrsim 5\times10^{-3}$.
However, in making this estimate, we assumed that any non-axisymmetric instability taking place deep inside the star strongly affects the observed surface field. 
This is not necessarily true, in which case higher $\Delta\Omega/\Omega_0$ would be compatible with the Auri\`ere scenario.

Given the present uncertainties on the stability condition (see previous sub-section), the level of the differential rotation and the effect of an internal instability on the surface field, 
the quantitative comparison between the theoretical critical field and the observed lower bound of Ap/Bp fields is still rather approximative. 
More realistic 3D simulations will help, although the density contrasts encountered in a stellar envelope remain beyond reach for such simulations.
It will be more relevant to compare the dependence of the critical field on the surface rotation rate with observations. Existing data indeed point towards a linear
dependency of the lower bound of Ap magnetic fields \citep{lignieres2014} compatible with the stability condition (\ref{condition_b0}).

\section{Conclusion}

We performed an extensive study of the evolution of a magnetic field in a differentially rotating radiative zone. 
Among other results, we provided a systematic estimate of the ratio between the toroidal and the poloidal magnetic field.
We tested the influence of the diffusivities and of the density profile and characterized the impact of different boundary conditions and differential rotation profiles.
The large parametric study that 2D numerical simulations made possible has allowed finding asymptotic regimes for the diffusion and the density contrast. 
This in turn enabled us to extrapolate our numerical results to realistic stellar diffusivities and density contrasts. 
A simple one-dimensional standing wave model was also found to provide reliable estimates of $max(B_\varphi/B_p)$ and the time at which this maximum is reached.

We then discussed the stability of the magnetic configurations dominated by the toroidal component using a local dispersion relation derived by \cite{ogilvie2007} and results of full 3D numerical simulations performed in a simplified case \citep{jouve2015}.
We argued that these magnetic configurations are likely to be subject to a magneto-rotational instability or to the Tayler instability, the nature of the instability depending solely on the ratio of the rotation rate to the toroidal Alfv\'en frequency $\Omega/\omega_{A_{\varphi}}$. 
Such non-axisymmetric instabilities are expected to modify the large-scale axisymmetric configuration when the surface poloidal field is weaker than a threshold value $B_{crit}$. 
In the context of the Auri\`ere scenario to explain the dichotomy of intermediate-mass star magnetism, our estimate of $B_{crit}$ is compatible with the lower bound of Ap/Bp magnetic fields, although a precise quantitative comparison remains difficult at this stage. 

We neglected meridional flows in our simulations.
The effect of a prescribed circulation has been considered by \cite{MMT88} and \cite{MMT90}. 
They found that it is negligible as long as the circulation velocity is very low compared to the Alfvén speed, which is the case for typical Eddington-Sweet-type circulations.
Another hypothesis in this work is that of axisymmetry, when it is known from observations that Ap/Bp stars are generally oblique rotators.
Non-axisymmetric components of the initial poloidal field could have an influence on the stability of the magnetic field.
\cite{MMT90}, \cite{moss92}, and \cite{WG2015} have shown that if $B_0/R \Omega_0\sqrt{4\pi\rho_0}\ll1$ the magnetic field is symmetrized before the differential rotation decays.
In such extreme cases, the magnetic configurations are predicted to be unstable by (\ref{condition_b0}).
In the opposite regime, $B_0/R \Omega_0\sqrt{4\pi\rho_0}\gg1$, the strong magnetic field quickly suppresses the differential rotation and the magnetic field remains inclined with respect to the rotation axis.
These situations are similar to axisymmetric cases in which magnetic configurations are also predicted to be stable.
In intermediate cases, we do not know the effect of the non-axisymmetric components of the field on the stability conditions.
This question needs to be addressed.

To proceed, we need 3D numerical simulations that include the effects of the stable stratification as well as the mechanism that forces the differential rotation. 
The pre-main-sequence contraction phase can in principle generate the required differential rotation \citep{lignieres2014}. 
While realistic diffusivities and density contrasts are beyond what can be reached by 3D simulations, we expect that the dependency of $B_{crit}$ with the rotation rate will not strongly depend on these parameters. 
The relation between $B_{crit}$ and the surface rotation will then be compared to observational constraints on the lower bound of Ap/Bp magnetic fields. 
Another application of such simulations is to consider the role of non-axisymmetric instabilities on the efficiency of the angular momentum transport in stellar radiative zones.

\begin{acknowledgements}
The authors acknowledge financial support from the Programme National de Physique Stellaire (PNPS) of CNRS/ INSU and from the Agence Nationale pour la Recherche (ANR) through the IMAGINE project.
T. G. is supported by the Special Priority Program 1488 PlanetMag of the German Science Foundation.
The authors wish to thank S\'ebastien Deheuvels for providing us with a density profile from a stellar evolutionary model.
\end{acknowledgements}

\bibliographystyle{aa.bst}
\bibliography{biblio} 

\end{document}